\documentclass[aps,prl,twocolumn,floatfix,showpacs]{revtex4}

\usepackage{bm,bbm,amsmath,amssymb,amsbsy,amsthm,graphicx,subfigure,color,txfonts,multirow}
\usepackage{amsfonts}
\usepackage{array}
 
\usepackage[utf8x]{inputenc}
\usepackage[greek,english]{babel}
\usepackage{teubner}

\usepackage[framemethod=tikz]{mdframed}
\usepackage[colorlinks]{hyperref}
 
\hypersetup{
	bookmarksnumbered,
	pdfstartview={FitH},
	citecolor={blue}, 
	linkcolor={red},
	urlcolor={black},
	pdfpagemode={UseOutlines}
}

 \newcommand{\tr}[1]{\text{Tr}}
\newcommand{\ket}[1]{|#1\rangle}
\newcommand{\bra}[1]{\langle#1|}

\begin{document}

\title{How difficult is it to prepare a quantum state?}

\author{Davide Girolami}
\email{davegirolami@gmail.com}
\affiliation{$\hbox{Los Alamos National Laboratory, Theoretical Division, P.O. Box 1663
Los Alamos, NM 87545, USA}$ \\
$\hbox{Kavli Institute for Theoretical Physics, UCSB, Santa Barbara, CA 93106, USA}$}

\begin{abstract}
Consider a quantum system prepared in an input state. One wants to drive it into a target state. Assuming classical states and operations as free resources, I identify a geometric cost function which quantifies the difficulty of the protocol in terms of how different it is from a classical process. The quantity determines a lower bound to the number of commuting  unitary transformations required to complete the task. I then discuss the link between the quantum character of a state preparation and the amount of coherence and quantum correlations that are created in the target state.   
 \end{abstract}
 
\date{\today}
 
   \maketitle
\noindent{\it Introduction --}  Quantum systems promise to outperform classical devices in information processing  protocols, if prepared in certain complex configurations \cite{nielsen}. 
It is then interesting to measure the difficulty  to drive a quantum system into a target state, and design the best strategies to complete the task.  Previous works determined time optimal Hamiltonian control dynamics \cite{carlini,caneva,cafaro,geonielsen,nielsen2,kha,allegra}, and energy efficient  out-of-equilibrium driving for classical and quantum systems \cite{schlogl,crooks,crooks2,deffner,deffner1}. Yet, 
 the difficulty of   preparing a  useful quantum state is not plainly due to the consumption of physical resources, as classical processes can take more time or energy than quantum ones. An alternative, informative metric should evaluate how different is a transformation  from  being classical. While measures of quantum coherence and correlations mark the difference  between classical and quantum states \cite{cohrev,modirev},  there is no clear boundary between classical and quantum processes, as there is no quantifier of their quantumness.  

\noindent Here I introduce a measure of  the  difficulty of a quantum state preparation in terms of how different it is from  a classical process. First,  I identify incoherent states and classical stochastic maps as well-motivated sets of free states and operations, being the only ones which do not display quantum superpositions, i.e. coherence. Creating coherence should be never easy because it can be sufficient for outmatching classical devices. This was proved by works in quantum information \cite{cohrev}, cryptography \cite{benn}, and communication \cite{asymrev}, which yet left  exact laws establishing how difficult is creating  superpositions  to be found.
Then, I introduce a design principle for  quantum driving  of general validity. The best preparation strategy is the input/target dynamics which minimizes a geometric index quantifying the quantum character of the transformation.  The geometric measure, which overcomes the limitations of customary distance functions,  lower bounds the operationally meaningful algorithmic cost  to prepare a state via commuting  operations. As a target state is expected to be computationally useful, it is also interesting to establish a link to the creation of quantum resources. I  derive  quantitative relations between the quantumness of a process, a computable lower bound, and the coherence and quantum correlations created in the target.  

\noindent{\it Quantum state preparation: free resources and cost  --} Suppose a finite dimensional quantum system is in a state described by a density matrix $\rho$. How hard is it to drive the system into a target state $\tau$? I formulate the problem in a geometric framework \cite{amari,geo}. The system dynamics is modeled  by a curve $\gamma: t\rightarrow \gamma_t$ in the stratified manifold of quantum states ${\cal M}$,  where $\gamma_t=\sum_{i}\lambda_{i}(t)\ket{i(t)}\bra{i(t)}, \sum_{i}\lambda_{i}(t)=1, \langle i(t)|j(t)\rangle=\delta_{i j}, \gamma_0\equiv\rho, \gamma_T\equiv \tau,$ is the spectral decomposition of the system state at time $t$. First, I identify what it is ``easy'' to obtain and to do. In the parlance of quantum information theory, this is represented by the free states and the free operations, respectively \cite{resource,resrev}.  I aim at associating the difficulty of the computation $\rho \rightarrow \tau$ with its quantumness. Thus, free state preparations must be classical processes, which are characterized as follows. If an input state $\rho=\sum_i \lambda_i(0)\ket{i(0)}\bra{i(0)},  \ket{i(0)}\equiv\ket{i_{\text{R}}}, \langle i_{\text{R}}\ket{j_{\text{R}}}=\delta_{ij},$ is given for free,  then any state which is diagonal in the reference basis $\{\ket{i_{\text{R}}}\}$ can be prepared (deterministically or stochastically) via an operation such that the state of the system is at any time described by an element of  ${\cal M}_{i_{\text{R}}}=\{ \tilde{\rho}=\sum_i \tilde{\lambda}_i\ket{i_{\text{R}}}\bra{i_{\text{R}}}\}$.
The information about the basis is then redundant and the transformation is at any time a classical  process. Hence, {\it the free states are the density matrices in} ${\cal M}_{i_{\text{R}}}$. {\it The free operations are the  maps  such that the state of the system  is at any time diagonal in a reference basis, $\gamma_t=\sum_{i} \lambda_{i} (t) \ket{i_{\text{R}}}\bra{i_{\text{R}}} \in {\cal M}_{i_{\text{R}}}, \forall t\in[0,T]$}. Note that the eigenspaces  $\{i_{\text{R}}\}$ are not necessarily of multiplicity one,  and a state can be free with respect to more than one basis. \\ 
 I discuss a few examples to justify these definitions.  
   A transformation between two orthogonal states $\ket{i}\rightarrow \ket{j}, \langle i|j\rangle=\delta_{ij},$  can be synthesized via a unitary operation, as well as by a   classical ``amplitude damping'' map $\gamma_t=(1 -t/T)\ket{i}\bra{i}+t/T\ket{j}\bra{j}$, in which the density matrix is diagonal at any time in a  basis with elements $\{i,j\}$. Hence, it is not necessarily quantum. Conversely,  non-commutativity between input and output density matrices implies that the process is quantum \cite{note}, as classical maps cannot create superpositions $\ket{i}\rightarrow a\ket{i}+b\ket{j},  a,b \in \mathbb{C}$.  The quantumness of a process is independent of the basis in which the states are written.  A transformation between commuting states displaying coherence in a basis, e.g. $a\ket{i}+b\ket{j}\rightarrow a\ket{j}-b\ket{i}$, always admits a classical implementation $\gamma_t=(1-t/T)\ket{+}\bra{+}+t/T\ket{-}\bra{-},  \ket{+}=a\ket{i}+ b\ket{j}, \ket{-}=a\ket{j}-b\ket{i}$.  One observes that the free operations in a resource theory are often characterized by the form of their Kraus operators \cite{resrev}, but this is generally not sufficient to signal the quantumness of a transformation.   A parametrized Kraus set for the amplitude damping is given by $K_1=\ket{+}\bra{+}+\sqrt{1-t/T}\ket{-}\bra{-}, K_2=\sqrt{t/T}\ket{+}\bra{-}$. Yet, the very same Kraus set transforms the input $\ket{i}$ into a non-commuting output.  The quantum character of  the continuous time evolution of a state is independent of reparametrizations of $t$. On this hand, continuous time classical maps seem  more appropriate free operations for state preparation than incoherent operations \cite{cohrev}.  For example, the unitary qubit transformation $e^{-i \sigma_y t}$ is  a quantum map at any time $t$, but it is a (strictly) incoherent operation with respect to the basis $\{0,1\}$  for $t=k \pi/2$ \cite{winter,yadinprx}, creating coherence otherwise. It is hard to  justify why a phase shift should be  easy only for some values of   $t$, as no experimental challenge emerges to implement this map  at different times.\\

\begin{figure}[t] 
\includegraphics[height=4.5cm,width=4.5cm]{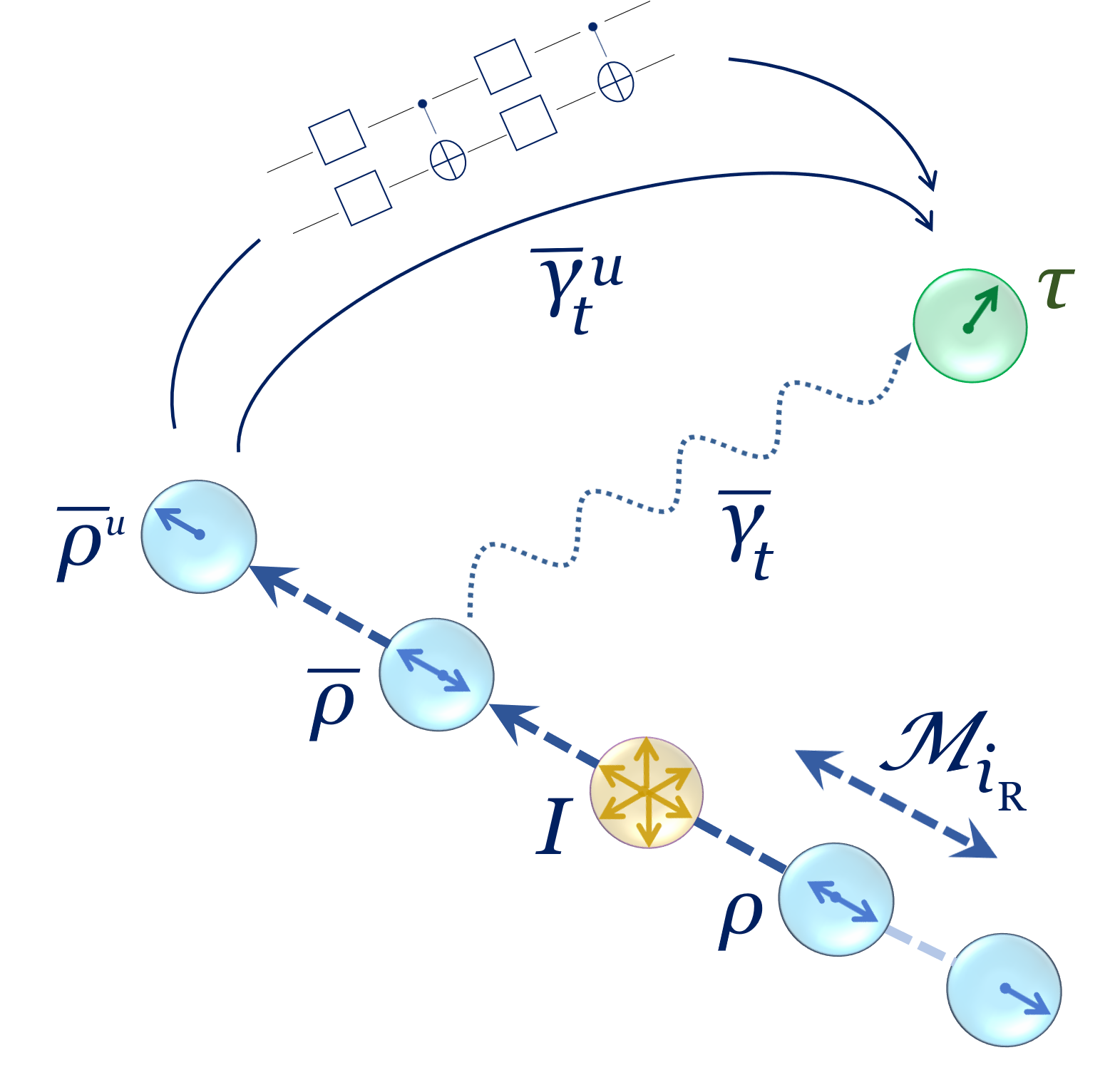}
\caption{The optimal path $\bar{\gamma}_t$ to drive a system from $\rho \in{\cal M}_{i_{\text{R}}}$ into $\tau$ is the minimizer of the quantum component of the energy $Q_{\rho}(\tau)=E_q^{\bar{\gamma}_t}(\bar{\rho}\rightarrow\tau)$ over all the free  states. The unitary map $\bar{\gamma}^u_t$ generates the energy minimizing path from a free state $\bar{\rho}^u$ isospectral to the target, $Q^u_{\rho}(\tau)=E^{\bar{\gamma}^u_t}(\bar{\rho}^u\rightarrow\tau)$.} 
\label{path}
\end{figure}

\noindent The difficulty of an input/target transformation can be then evaluated in terms of how different it is from a free operation, i.e. a classical process. This cannot be measured by  distance functions,  which quantify the ability to distinguish two states via measurements \cite{nielsen,wootters}. For example, two orthogonal  states $\ket{i}, \ket j,$ are more distinguishable than  $\ket{i}$ and any state displaying coherence $a\ket{i}+b\ket{j}, a,b \neq 0$. 
 I search for a function of  input   and target states $Q_\rho(\tau)$ which meets a set of desirable properties: faithfulness, being zero only when the target is a free state, $Q_\rho(\tau)=0\Leftrightarrow \tau\in {\cal M}_{i_{\text{R}}}$; invariance under free operations, taking the same value for all free states, $Q_\rho(\tau)=Q_{\tilde{\rho}}(\tau), \forall \tilde{\rho}\in {\cal M}_{i_{\text{R}}}$; contractivity under  mixing,  $Q_\rho(\tau)\geq Q_{\Gamma(\rho)}(\Gamma(\tau))$, where $\Gamma$ is a completely positive trace-preserving (CPTP) map.  \\
 Consider the energy of a  curve at fixed boundaries 
\begin{eqnarray}\label{energy}
E^{\gamma_t}(\rho\rightarrow \tau):=\int_{0}^{T} ||\dot{\gamma}_t||^2 d t, \gamma_0\equiv \rho, \gamma_T\equiv \tau,
 \end{eqnarray}
where the norm is induced by a Riemannian Fisher metric, the only class of contractive metrics under noisy maps on ${\cal M}$ \cite{petz,moro,luoskew}.
The quantity is formally equivalent to  the kinetic energy (per unit of time)  for a particle traveling on the manifold \cite{petersen,tao,luokin}, while being generally not related to the physical energy. By decomposing the state as $\gamma_t= U_t \Lambda_t U_t^\dagger, U_0=I,$ where $\Lambda_t$ is a diagonal matrix with the state eigenvalues as entries,  
the rate of change reads $\dot{\gamma}_t=U_t\dot{\Lambda}_tU_t^\dagger+i[\gamma_t,H_t], H_t=i \dot{U}_tU_t^\dagger$.   For classical processes,  only the first term survives at any time $t$. 
 On the other hand,  a unitary transformation  $\gamma^u_t=U_t\Lambda_0U_t^\dagger, \forall t$, is  genuinely quantum. It changes the state eigenbasis while the spectrum is invariant, so only the second term appears  at any time $t$. 
  For a path corresponding to a general CPTP map, the two terms coexist. 
The key point is that, independently of the specific metric employed, the tangent space to  $\mathcal{M}$ has a direct sum structure such that $||\dot{\gamma}_t||^2=||U_t\dot{\Lambda}_tU_t^\dagger||^2+||i[\gamma_t,H_t]||^2$  \cite{geo,amari} .  Hence, it is possible to discriminate between  classical and quantum components of the energy:
\begin{eqnarray}
 E^{\gamma_t}(\rho\rightarrow \tau)&=&E_c^{\gamma_t}(\rho\rightarrow \tau)+E_q^{\gamma_t}(\rho\rightarrow \tau),\\
 E_c^{\gamma_t}(\rho\rightarrow \tau)&:=&\int_{0}^{T} ||U_t\dot{\Lambda}_tU_t^\dagger||^2 d t, \nonumber\\ 
 E_q^{\gamma_t}(\rho\rightarrow \tau)&:=&\int_{0}^{T} ||i[\gamma_t,H_t]||^2 d t.\nonumber
\end{eqnarray}
Note that a distance function cannot be split. For unitary transformations, only the quantum term survives, capturing the sensitivity of the system to phase shifts. This property, called asymmetry \cite{asymrev}, is the peculiar resource for phase estimation. Generalizing the concept of asymmetry to arbitrary CPTP maps, the basis changing component of the (squared) speed  measures the sensitivity of the system in a state $\gamma_t$ to a map $\Gamma_t$ due to {\it quantum} effects. 
 Hence, the quantumness of a computation $\rho\rightarrow \tau$, i.e. the difficulty of driving the system into the target state within a time $T$, is given by the minimum quantum component of the energy over all the possible maps linking a free state to the target:
   \begin{eqnarray}\label{defq}
   Q_\rho(\tau):&=& E_q^{\bar{\gamma}_t}(\bar{\rho}\rightarrow\tau),\nonumber\\
   E_q^{\bar{\gamma}_t}(\bar{\rho}\rightarrow\tau)&=&\min\limits_{\tilde{\gamma}_t, \tilde{\rho}}E_q^{\tilde{\gamma}_t}(\tilde{\rho}\rightarrow\tau),\ \ \  \tilde{\rho}\in {\cal M}_{i_{\text{R}}}, \tilde{\gamma}_t:\tilde{\rho}\mapsto \tau.
    \end{eqnarray}
 The results I am going to present would hold for any Riemannian metric. Yet, for the sake of clarity, I  employ from now on  the Bures metric, which plays an important role in quantum statistics and quantum information theory \cite{petz,helstrom,metrorev,nielsen}. 
 The squared speed of the system at time $t$ is
 \begin{eqnarray}
 ||\dot{\gamma}_t||^2=\sum_i \frac{\dot{ \lambda_i}(t)^2}{4\lambda_i(t)}  
 + \sum_{i< j} \frac{|\langle i(t)|i[\gamma_t,H_t]|j(t)\rangle|^2}{\lambda_i(t)+\lambda_j(t)}. \end{eqnarray}
   The first term is the squared norm related to the classical Fisher metric, while the second one is the quantum contribution. For unitary transformations $\gamma^u_t$, only the second term survives, $E_q^{\gamma^u_t}(\rho\rightarrow \tau)=E^{\gamma^u_t}(\rho\rightarrow \tau)$ \cite{diogo,taddei}. This is non-negative, vanishing at any time only for classical processes,  and non-increasing under mixing \cite{notefisher,benfisher}. If the evolution is time-independent, $U_t=e^{-i Ht}$, the quantity is lower bounded by  $T$ times the variance of  the Hamiltonian,  $E^{\gamma^u_t}(\rho\rightarrow \tau)\leq T V_{\rho}(H), V_{\rho}(H):=\text{Tr}(\rho H^2)-\text{Tr}(\rho H)^2$, being the inequality   saturated for pure states.   
 It  follows from the properties of the quantum Fisher information, i.e. the instantaneous (squared) speed,  that the required constraints are met. Faithfulness holds because if and only if the target is a free state, there exists a classical   preparation such that $||\dot{\gamma}_t||^2=\sum_i \frac{\dot{ \lambda_i}(t)^2}{4\lambda_i(t)}, \forall t$. Invariance under free transformations of the input state is satisfied by construction. Defining $\Gamma(\gamma_t): \Gamma(\rho)\mapsto\Gamma(\tau)$ the dynamics of a state subject at any time to a CPTP map, the quantity is contractive, $Q_\rho(\tau)\geq E_q^{\Gamma(\bar{\gamma}_t)}(\Gamma(\rho)\rightarrow\Gamma(\tau))\geq Q_{\Gamma(\rho)}(\Gamma(\tau))$.\\ 
The definition in Eq.~\ref{defq} unrealistically  assumes that every dynamics linking input and target states is implementable in practice. I therefore derive an operationally motivated upper bound (see Fig.~\ref{path}). Suppose only   classical processes and unitary transformations are allowed. This is not very limiting: Any preparation can be split into a change of spectrum and a change of basis, $\rho\rightarrow\rho^u\rightarrow\tau, \rho^u=\sum_i\lambda_i(T)\ket{i_{\text{R}}}\bra{i_{\text{R}}}\in {\cal M}_{i_{\text{R}}},$ where $\lambda_i(T)$ are the eigenvalues of $\tau$. The first step can be completed via a free operation.  The second step can be completed via one purely quantum, unitary change of basis $\gamma_t^u: \rho^u\mapsto \tau$. One then has   $E^{\gamma^u_t}(\rho^u\rightarrow\tau)=E_q^{\gamma^u_t}(\rho^u\rightarrow\tau)$.  
For a target state of a $d$-dimensional system with eigenvalues having multiplicities $m_i$, there are $d!/(\Pi_i m_i!)$ isospectral free states  which can freely transform into each other via  permutations, $\rho^u_p=P\rho^u P^\dagger$.
 The minimum energy to complete the second step is   computed by minimizing over the free states which are isospectral to the target. Thus, the difficulty to complete a state preparation with classical operations and unitaries is
\begin{eqnarray}
Q^u_\rho(\tau)&:=&E^{\bar{\gamma}^u_{t}}(\bar{\rho}^u\rightarrow\tau),\\
E^{\bar{\gamma}^u_{t}}(\bar{\rho}^u\rightarrow\tau)&=&\min\limits_{\gamma^u_{p,t}, \rho_p^u}E^{\gamma^u_{p,t}}(\rho_p^u\rightarrow\tau), \gamma^u_{p,t}:\rho_p^u\mapsto \tau.\nonumber
\end{eqnarray}    
\noindent One has $Q^u_\rho(\tau)\geq Q_\rho(\tau)$. This upper bound also meets by construction  faithfulness, invariance and contractivity, $Q^u_\rho(\tau)=0\Leftrightarrow \tau\in{\cal M}_{i_{\text{R}}}, Q^u_\rho(\tau)= Q^u_{\tilde{\rho}}(\tau), \forall \tilde{\rho}\in{\cal M}_{i_{\text{R}}},  Q^u_\rho(\tau)\geq Q^u_{\Gamma(\rho)}(\Gamma(\tau))$. Note that the two-step, classical-quantum split is optimal by construction.  A classical map is, by definition, a transformation in ${\cal M}_{i_\text{R}}$. Hence, the path corresponding to an arbitrary sequence of multiple classical and quantum steps returns to ${\cal M}_{i_\text{R}}$ multiple times, requiring more energy. \\

\noindent{\it Optimal path, algorithmic complexity and quantum resources -- } An important question is what is the best path $\bar{\gamma}^u_t$ to reach the target from an isospectral free state. The  map between two states which minimizes the energy $E^{\gamma_t}(\rho\rightarrow\tau)$ is the length minimizer at constant speed \cite{petersen}.
A distance function is $D(\rho,\tau):=\min\limits_{\gamma_t:\rho\mapsto\tau}\int_{0}^{T} ||\dot{\gamma}_t || dt$.
The one related to the Bures metric is the Bures angle $D_\text{B}(\rho,\tau)=\cos^{-1}\text{Tr}\left(|\sqrt{\rho}\sqrt{\tau}|\right)$.
 The energy minimizing map from a pure free state $\rho^u=\ket{\psi_{\rho^u}}\bra{\psi_{\rho^u}}$ to a pure target $\tau=\ket{\psi_{\tau}}\bra{\psi_{\tau}}$ is the length minimizing unitary, and the Bures angle reduces  to the Fubini-Study distance $D_{\text{FS}}(\psi_{\rho^u}, \psi_\tau)=\cos^{-1}|\langle\psi_{\rho^u}| \psi_\tau\rangle|$. The closest free pure state $\bar{\rho}^u$ to the target is then the one with maximal overlap. The  length/energy minimizing constant speed path reads 
 \begin{eqnarray}
 \bar{\gamma}^u_t&=&\ket{\psi_{\bar{\gamma}_t^u}}\bra{\psi_{\bar{\gamma}^u_t}},\\ \ket{\psi_{\bar{\gamma}_t^u}}&=&(\cos{\theta}-\sin{\theta}/\tan{d})\ket{\psi_{\bar{\rho}^u}}+ (\sin{\theta}/\sin{d})\ket{\psi_{\tau}}, \nonumber
 \end{eqnarray}
 where $\theta=d\ t/T, d:=D_{\text{FS}}(\psi_{\bar{\rho}^u},\psi_{\tau}).$
  This is obtained by the expression for the length minimizing path \cite{uhl,ericsson,barnum}, and  noting that the energy minimizer is unique up to affine reparametrizations $t'=a t+b, a,b\in \mathbb{R}$.  Finding the optimal unitary for   mixed target states is more challenging, while necessary conditions for the shortest unitary path between isospectral states have been found \cite{andersson}. However,  the result  for pure states yields a lower bound to $Q^u_\rho(\tau)$ for arbitrary target states. The distance between two mixed states is the minimum distance between their purifications \cite{uhl}. The closest isospectral free state to the target is then the one with the closest purification $\ket{\psi^{\text{{\tiny purif}}}_{\bar{\rho}^u}}$ to a target purification   $\ket{\psi^{\text{{\tiny purif}}}_{\tau}}$. The  closest purifications have a compact expression \cite{barnum}, which in this case is 
 \begin{eqnarray}
  \ket{\psi^{\text{{\tiny purif}}}_{\bar{\rho}^u}}&=& \sum_i \sqrt{\bar{\rho}^u}\ket{i_{\text{R}}}\otimes \ket{i_{\text{R}}},\\
 \ket{\psi^{\text{{\tiny purif}}}_{\tau}}&=&\sum_i 1/\sqrt{\bar{\rho}^u} \sqrt{\sqrt{\bar{\rho}^u}\tau\sqrt{\bar{\rho}^u}} \ket{i_{\text{R}}}\otimes \ket{i_{\text{R}}}.\nonumber
  \end{eqnarray}
    The length/energy minimizing (generally not unitary) path  between two mixed states is obtained by partial trace along the shortest  (unitary) path  between the closest purifications. Thus, one has $Q_\rho^u(\tau)\geq Q_{\psi_{\bar{\rho}^u}^{\text{{\tiny purif}}}}(\psi^{\text{{\tiny purif}}}_{\tau})$. 
	 The inequality is saturated for pure targets. Such a lower bound, which satisfies by construction faithfulness, invariance and monotonicity properties, is computed as follows. Consider, for example, driving a qubit from an input state with Bloch form $\rho=1/2(I+q_z\sigma_z), q_z\neq 0,$ to a target $\tau=1/2(I+\vec{r}\cdot \vec{\sigma})$. The isospectral free states to the target are identified by $|q_z|= |\vec{r}|$.  One has $Q_{\psi^{\text{{\tiny purif}}}_{\bar{\rho}^u}}(\psi_{\tau}^{\text{{\tiny purif}}})=\left\{\cos^{-1}\left[\left(\sqrt{f_{+}}+\sqrt{f_-}\right)/2\right]\right\}^2/T, f_{\pm}= 1+|\vec{r}| r_z \pm \sqrt{|\vec{r}|^2+ 2 |\vec{r}| r_z -|\vec{r}|^2(|\vec{r}|^2-r_z^2-1)}$. The process is classical for $r_x=r_y=0, |\vec{r}|=|r_z|,$ while the maximum energy $\pi^2/(16 T)$ is required to prepare the pure state given by $ r_x^2+r_y^2=1,r_z=0$. The same method applies for systems of dimension $d>2$, as their states admit a Bloch form $1/d(I+\vec{r}\cdot\vec{\Sigma}), $ where  $\vec{\Sigma}$ is a vector of  $d\times d$ traceless matrices.\\

\noindent The geometric index  $Q^u_\rho(\tau)$ can bound the size of  preparation algorithms. Suppose that a unitary map $\gamma^u_t=\rho^u\mapsto\tau$  is synthesized by  $
N$  commuting unitary operations, e.g. logic gates, $\gamma^u_t=U_{t}\rho^u U^\dagger_t, U_t=e^{-i H t }, H=\sum_{l=1}^NH_l, [H_l,H_k]=0, \forall l,k$. The scenario describes the phase imprinting step in parallel estimation protocols \cite{metrorev}, and the preparation of highly entangled symmetric states, $(a\ket{0}+b\ket{1})\otimes \ket{0}^{\otimes N}\rightarrow a\ket{0}^{\otimes N+1}+b\ket{1}^{\otimes N+1}, a,b \in \mathbb{C},$ via controlled gates between the first and th $l+1$-th qubit. Consider the seminorm of each Hamiltonian  $|H_l|=h_{l,M}-h_{l,m}$ being  the difference between its largest and smallest eigenvalues  \cite{geremia}.  It measures the  complexity of $H_l$, as it depends on the number of gates  implementing the Hamiltonian, and the size of the correlations they can build \cite{geremia,kok}.
Since  $4V_{\rho}(H)\leq|H|^2\leq (\sum_l |H_l|)^2$, one has $E^{\gamma^u_t}(\rho\rightarrow\tau)  \leq   T N^2 \overline{|H|^2}/4,$ 
 where $\overline{|H|^2}$ is the average  squared seminorm over all the generators $H_l$. By assuming  every  Hamiltonian to have the same seminorm $|H_l|=h,\forall l$, one has
 \begin{eqnarray}
 N\geq \frac{2}{h}\left(\frac{Q^u_\rho(\tau)}{T}\right)^{1/2}.
  \end{eqnarray}
   The bound is saturated for superpositions of the largest and smallest eigenvalues of $H$, $\ket{\psi_{\rho}}=(\ket{h_M}+e^{i\phi}\ket{h_m})/\sqrt2$, which are the most sensitive inputs to the map.\\
   
\noindent The quantumness of a transformation  is also linked to the coherence the target displays with respect to the reference basis \cite{herbut,bau,cohrev}, here quantified by the distance to the set of incoherent states ${\cal C}^{i_{\text{R}}}_{\text{B}}(\tau):=\min\limits_{\tilde{\rho}\in{\cal M}_{i_{\text{R}}}}D_{\text{B}}(\tilde{\rho},\tau)$. (The distance function is determined by the chosen Riemannian metric.) One has 
\begin{eqnarray}
 Q^u_\rho(\tau)\geq Q_{\psi^{\text{{\tiny purif}}}_{\bar{\rho}^u}}(\psi_{\tau}^{\text{{\tiny purif}}}) \geq \left({\cal C}^{i_{\text{R}}}_{\text{B}}(\tau)\right)^2/T \geq Q_\rho(\tau),
 \end{eqnarray} 
 where $Q^{(u)}_\rho(\tau)=0\Leftrightarrow  {\cal C}^{i_{\text{R}}}_{\text{B}}(\tau)=0$. The chain holds as  $Q_{\psi^{\text{{\tiny purif}}}_{\bar{\rho}^u}}(\psi_{\tau}^{\text{{\tiny purif}}})=D^2_{\text{B}}(\bar{\rho}^u,\tau)/T$.
For pure states, one has ${\cal C}^{i_{\text{R}}}_{\text{B}}(\psi_\tau)=\cos^{-1}\max\limits_{i_{\text{R}}}|\langle i_{\text{R}}|\psi_{\tau}\rangle|$ \cite{geobound}, which implies $
D_{\text{B}}(\psi_{\bar{\rho}^u},\psi_\tau)={\cal C}^{i_{\text{R}}}_{\text{B}}(\psi_\tau)$. In the multipartite case, the quantumness of the transformation upper bounds the  quantum correlations in the target, whenever the reference basis is local or multi-local. Note that rather than the average/maximum ability of a map to create quantumness \cite{zanardi,meznaric}, I compute the minimum cost. The most general form of bipartite quantum correlations, quantum discord \cite{modirev}, can be measured by the minimum coherence over all the bi-local bases, ${\cal D}_{\text{B}}(\rho_{12}):=\min\limits_{i_{1}j_{2}}{\cal C}^{i_{1}j_{2}}_{\text{B}}(\rho_{12}), \{i_{1}j_{2}:=\ket{i_1}\otimes\ket{j_2}\}$. This is  the symmetric discord \cite{luo,amid}, but the argument applies to the original asymmetric definition as well.   Consider the set of free states being the zero discord states $\rho_{12}=\sum_{ij}p^{12}_{ij}\ket{i_{1}j_{2}}\bra{i_{1}j_{2}}, \sum_{ij}p^{12}_{ij}=1, \rho_{12}\in {\cal M}_{i_{1}j_{2}}$. That is, the reference basis is the bi-local basis $\{i_{1}j_{2}\}$. One has 
\begin{eqnarray}
  Q_{\psi^{\text{{\tiny purif}}}_{\bar{\rho}^u\in  {\cal M}_{i_{1}j_{2}}}}(\psi_{\tau_{12}}^{\text{{\tiny purif}}})\geq {\cal D}^2_{\text{B}}(\tau_{12})/T. 
\end{eqnarray}
Yet,  ${\cal D}_{\text{B}}(\tau_{12})=0 \nRightarrow Q^{(u)}_{\rho\in  {\cal M}_{i_{1}i_{2}}}(\tau_{12})=0$.
  For example, the qutrit-qubit map $p\ket{0}\bra{0}\otimes\ket{0}\bra{0}+(1-p)\ket{1}\bra{1}\otimes\ket{1}\bra{1}\rightarrow p\ket{0}\bra{0}\otimes\ket{0}\bra{0}+(1-p)/2[(\ket{1}+\ket{2})(\bra{1}+\bra{2})\otimes\ket{1}\bra{1}]$ does not create discord, but it generates coherence with respect to the basis $\{0,1,2\}$ \cite{yadinprx}.
  I extend the bound to 
 an hierarchy of measures of coherence and genuine multipartite correlations of different orders \cite{multicoh1,multicoh2,giorgi,weaving,chinamulti}. Given an $N$-local reference basis $\{i_{1}\dots i_{N}\}$,  the coarse grained bases containing up to $k$-local terms read $\{i^k\}:=\{i_{12\ldots k_1}i_{k_1+1 k_1+2\ldots k_2}\ldots i_{k_{j-1}+1 k_{j-1}+2\ldots k_j}\}, \sum_j k_j=N, k\geq k_j, \forall j$.  The Bures quantum discord of ``order higher than $k$'' in an $N$-partite target state $\tau_{1\ldots N}$ is ${\cal D}_{\text{B}}^{k\rightarrow N}(\tau_{1\ldots N}):=\min_{i^k} {\cal C}^{i^k}_{\text{B}}(\tau_{1\ldots N})$. Suppose the free states  to be the incoherent states in a coarse grained basis $\bar{i}_k$,  $\rho_{1\ldots N}=\sum_{\bar{i}^k}p_{\bar{i}^k}\ket{\bar{i}^k}\bra{\bar{i}^k}$, i.e. a subset of the states without quantum discord of order higher than $k$. One has 
 \begin{eqnarray}
 Q_{\psi^{\text{{\tiny purif}}}_{\bar{\rho} \in {\cal M}_{\bar{i}^k}}}(\psi_{\tau_{1\ldots N}}^{\text{{\tiny purif}}})\geq  \left({\cal D}_{\text{B}}^{k\rightarrow N}(\tau_{1\ldots N})\right)^2/T.
 \end{eqnarray}
   A bound for the total amount of quantum correlations is obtained for $k=1$, where ${\cal D}_{\text{B}}^{1\rightarrow N}(\tau_{1\ldots N})$ is the distance of the target to the  classically correlated  states. 
 As   geometric measures of discord   upper bound   measures of entanglement \cite{modigeo}, being equal to them for pure states, the quantumness of a process   upper bounds  measures of multipartite entanglement  in the target. \\ 
    
\noindent{\it Conclusion --} I have quantified the difficulty of preparing a quantum system in a target state 
by measuring the process quantumness. The optimal dynamics is obtained by solving the geometric problem of minimizing the quantum contribution to the energy of the associated curve. The result highlights the usefulness of geometric methods to establish fundamental limits of quantum information processing. 
Geometric bounds could provide a benchmark to evaluate the performance of methods for shortening quantum algorithms \cite{auto}, which is of renewed interest due to the applicability of machine learning techniques.  Also, the resource theory approach can be fruitful to solve critical quantum control problems \cite{quaint}. \\

 \noindent{\it Acknowledgements. --}
I thank F. Anz\`{a}, I. Bengtsson, C. Cafaro, P. Gibilisco, A. Jencova, S. Luo, R. Maity, V. Moretti, B. Yadin and K. Zyczkowski for fruitful discussions. I acknowledge support from LANL through the LDRD project 20180702PRD1 and the LDRD Rapid Response project ``Unraveling Entanglement in a black box quantum computer''. Also, this research was supported in part by the National Science Foundation under Grant No. NSF PHY-1748958,  as part of the work was carried out at the KITP in S. Barbara.


\begin{thebibliography}{99}
\bibitem{note}A transformation between commuting states may not be free.  For example, $\ket{00}\rightarrow(\ket{01}+\ket{10})/\sqrt{2}$ is not free if the reference basis is $\{\ket{0,1}\}^{\otimes 2}$, while it is free if the basis is $\{\ket{00}, \ket{11},  (\ket{01}\pm\ket{10})/\sqrt2\}$. In the second case, entanglement cannot be detected via free operations.
\end{thebibliography}
    \end{document}